\documentclass[%
 reprint,
 amsmath,amssymb,
 aps,
]{revtex4-1}

\usepackage{graphicx}
\usepackage{dcolumn}
\usepackage{bm}
\usepackage{soul}
\usepackage{physics}
\usepackage{xcolor}
\usepackage{mathrsfs}
\usepackage{dsfont}

\setstcolor{red}

\def\bfr{{\bf r}}

\def\bfx{{\bf x}}
\def\bfy{{\bf y}}

\def\bfA{{\bf A}}
\def\bfB{{\bf B}}
\def\bfE{{\bf E}}

\def\bfI{{\bf I}}

\def\ssB{{\scriptscriptstyle B}}

\def\ssF{{\scriptscriptstyle F}}

\def\ssN{{\scriptscriptstyle N}}

\def\\mfK \mfK {{\scriptscriptstyle \mfK \mfK }}

\def\c\mfK {\mathcal{\mfK }}

\def\cO{\mathcal{O}}

\def\cS{\mathcal{S}}

\def\mff{\mathfrak{f}}
\def\mfg{\mathfrak{g}}

\def\mfK{\mathfrak{K }}

\def\msC{\mathscr{C}}
\def\msD{\mathscr{D}}

\def\exx{{\mathfrak{s}}}
\def\QED{{\scriptscriptstyle QED}}
\def\NS{{\scriptscriptstyle NS}}

\newbox\charbox
\newbox\slabox
\def\slsh#1{{      
        \setbox\charbox=\hbox{$#1$}
        \setbox\slabox=\hbox{$/$}
        \dimen\charbox=\ht\slabox
        \advance\dimen\charbox by -\dp\slabox
        \advance\dimen\charbox by -\ht\charbox
        \advance\dimen\charbox by \dp\charbox
        \divide\dimen\charbox by 2
        \raise-\dimen\charbox\hbox to \wd\charbox{\hss/\hss}
        \llap{$#1$}
}}

\def\exd{{\hbox{d}}}

\def\bfx{{\bf x}}
\def\bfy{{\bf y}}

\def\bfr{{\bf r}}

\def\bfE{{\bf E}}
\def\bfA{{\bf A}}
\def\bfB{{\bf B}}

\def\cO{{\cal O}}

\def\pref#1{(\ref{#1})}
\def\nn{\nonumber}
\def\bea{\begin{eqnarray}}
\def\eea{\end{eqnarray}}
\def\be{\begin{equation}}
\def\ee{\end{equation}}

\def\exd{{\rm d}}

\def\ol#1{\overline{#1}}

\def\ssB{{\scriptscriptstyle B}}

\def\ssF{{\scriptscriptstyle F}}

\def\ssN{{\scriptscriptstyle N}}

\def\nn{\nonumber}

\def\({\left(}
\def\){\right)}

\def\pref#1{(\ref{#1})}

\begin{document}

\preprint{Preprint Number}

\title{Nuclear Predictions for $H$ Spectroscopy without Nuclear Errors}

\author{C.P.~Burgess}
\author{P.~Hayman}%
\author{Markus Rummel}
\author{L\'aszl\'o Zalav\'ari}
\affiliation{%
 Physics \& Astronomy, McMaster University, Hamilton, ON, Canada, L8S 4M1\\
 Perimeter Institute for Theoretical Physics, Waterloo, Ontario N2L 2Y5, Canada
}%

\date{\today}

\begin{abstract}
Nuclear-structure effects often provide an irreducible theory error that prevents using precision atomic measurements to test fundamental theory. We apply newly developed effective field theory tools to Hydrogen atoms, and use them to show that (to the accuracy of present measurements) all nuclear finite-size effects ({\it e.g.}~the charge radius, Friar moments, nuclear polarizabilities, recoil corrections, Zemach moments {\it etc.}) only enter into atomic energies through exactly two parameters, independent of any nuclear-modelling uncertainties. Since precise measurements are available for more than two atomic levels in Hydrogen, this observation allows the use of precision atomic measurements to eliminate the theory error associated with nuclear matrix elements. We apply this reasoning to the seven atomic measurements whose experimental accuracy is smaller than 10 kHz to provide predictions for nuclear-size effects whose theoretical accuracy is not subject to nuclear-modelling uncertainties and so are much smaller than 1 kHz. Furthermore, the accuracy of these predictions can improve as atomic measurements improve, allowing precision fundamental tests to become possible well below the `irreducible' error floor of nuclear theory.

\end{abstract}

\pacs{Valid PACS appear here}
\maketitle


\section{\label{sec:intro}Introduction}

The comparison of precision measurements and detailed calculations for transition frequencies in Hydrogen-like atoms have taught us much about fundamental physics, going back to the discovery of quantum mechanics itself. 

More recently, modern experimental techniques have pushed the accuracy of these measurements to ever-higher precision, and one might hope that by bringing ever-smaller effects into experimental reach this improvement should provide ever-stronger insights about physics at sub-nuclear scales. Unfortunately this program has run into obstacles, largely because of the sad fact that atomic nuclei are complicated objects that are governed by strong interactions that resist precision {\it ab initio} calculations of nuclear properties. Uncertainties in calculating how nuclear structure affects atomic energies put a floor on the size of theoretical errors in atomic physics, a floor larger than many of the small effects whose measurement is sought.  

In this paper we aim to push past this floor in theoretical error by systematically identifying combinations of energy differences from which all nuclear-physics uncertainties cancel (with more details given in \cite{PPEFT-Hyd}, and extending earlier arguments for spinless nuclei \cite{ppeftA}). We do so by following standard practice that computes atomic energies as a triple expansion in powers of $R/a_\ssB = mRZ\alpha$, $m/M$ and $\alpha$ around the familiar lowest-order Bohr formula (we use units with $\hbar = c = 1$)
\be
  \varepsilon_n = - \frac{(Z\alpha)^2m_r }{2n^2} \,,
\ee
where $m_r = mM/(m+M) = m + \cO(m^2/M)$ is the reduced mass, with $m$ and $M$ the electron and nuclear masses, while $n = 1,2,\cdots$ is the usual principal quantum number. (It is useful to keep the nuclear charge $Z$ as a variable when calculating, though we take $Z = 1$ in final applications.) 

Our main result is this: by exploiting the fact that the $R/a_\ssB$ expansion is a low-energy approximation and so is most efficiently organized using effective field theory (EFT) methods, we show that nuclear structure can only enter into atomic energy shifts through a total of $N_{\rm nuc} = 2$ parameters, at least working out to (and including) order $m^3R^2(Z\alpha)^6$ and $m^4 R^3 (Z\alpha)^5$. This result might seem a surprise given that explicit nuclear calculations indicate that more than two independent nuclear moments appear to contribute at this order, including the Friar, Zemach and related moments, nuclear polarizabilities, and so on. 

We show that an EFT treatment reveals that all these nuclear moments only contribute to atomic energy shifts through two independent parameters, ultimately because of the exceedingly low energies with which electrons probe nuclei. Although EFT methods are not new to this type of atomic analysis \cite{boundNRQED}, what is more recent is the development of first-quantized EFT methods \cite{ppeft1, ppeft2, ppeft3, falltocenter}, in terms of which the matching of effective couplings is expressed as boundary conditions. This provides a more efficient counting of the ways effective nuclear couplings can enter into atomic observables, and it is the exploitation of this that underlies our calculations. Our main results are the energy-shift formulae \pref{ppeftAenergy+} through \pref{ppeftAenergy-}, which agree well with standard expressions in terms of nuclear moments \cite{friar, zemach, eides, pachucki2018, kalinowski2018} once the two parameters in these formulae are given by \pref{eepsstarQED} and \pref{eepsFQED}. More details and wider applications are provided in \cite{PPEFT-Hyd}. 

The observation that general EFT principles only allow nuclear structure to contribute through $N_{\rm nuc} = 2$ independent parameters means that nuclear physics can be completely sidestepped whenever there are $N_{\rm exp} > 2$ well-measured transitions involving a specific type of atom. While this observation is not remarkable in itself, what is surprising is how small $N_{\rm nuc}$ turns out to be. Inferring the values of the two independent parameters from atomic observations allows nuclear contributions to all other transitions to be computed without the need to rely on nuclear models, and their associated uncertainties. Examples of such predictions for atomic Hydrogen can be found in Table \ref{ahprecise}, where the above accuracy translates into nuclear errors of order $10^{-2}$ kHz.   

\begin{table*}[t]
\centering
\begin{tabular}{|c|c|c|c|c|c|}
\hline
Transition & $\nu_{\rm exp}$ &  $E_{\rm nuc}$ & $\Delta E_{\rm exp}$ & $\Delta E_{\rm th}$ & $\Delta E_{\rm trunc}$\\ [3pt]
\hline
$2P_{1/2}^{\ssF=1} - 2S_{1/2}^{\ssF=0}$ & 909  871.7(3.2)  & $-143.70$ & 0.0069 & 0.57 & 0.00031 \\ [5pt]
\hline
$1S_{1/2}^{\ssF=1} - 1S_{1/2}^{\ssF=0}$ & 1 420  405.751 768(1)  & $-57.75$ & 0.054 & 4.5 & 0.0033 \\ [5pt]
\hline
$8S_{1/2}^{\ssF=1} - 2S_{1/2}^{\ssF=1}$ & 770 649  350  012(9)  & $-134.348$ & 0.0014 & 0.080 & 0.00010\\ [5pt]
\hline
$8D_{3/2}^{\ssF=2, 1} - 2S_{1/2}^{\ssF=1}$ & 770  649 504  450(8)  & $-136.481$ & 0.0014 & 0.081 & 0.00010 \\ [5pt]
\hline
$8D_{5/2}^{\ssF=3, 2} - 2S_{1/2}^{\ssF=1}$ & 770  649  561  584(6)  & $-136.481$ & 0.0014 & 0.081 & 0.00010\\ [5pt]
\hline
$12D_{3/2}^{\ssF=2, 1} - 2S_{1/2}^{\ssF=1}$ & 799  191  710  473(9)  & $-136.481$ & 0.0014 & 0.081 & 0.00010\\ [5pt]
\hline
$12D_{5/2}^{\ssF=3, 2} - 2S_{1/2}^{\ssF=1}$ & 799 191  727  404(7)  & $-136.481$ & 0.0014 & 0.081 & 0.00010\\ [5pt]
\hline
$3S_{1/2}^{\ssF=1} - 1S_{1/2}^{\ssF=1}$ & 2 922 743 278 671.5 (2.6)  & $-1051.35$ & 0.011 & 0.62 & 0.00079 \\ [5pt]
\hline
\end{tabular}
\caption{Transitions from  \cite{hessels2019} (row 2), \cite{kramida} (rows 3-8) and from \cite{fleurbaey} (row 9) that are measured with better than 10 kHz accuracy in atomic Hydrogen. Column 2 gives their experimental values (and experimental errors in brackets); with all values given in kHz. Column 3 gives the nuclear-finite-size contribution to the transition energy predicted by eqs.~\pref{ppeftAenergy+}, \pref{spinenergy} and  \pref{ppeftAenergy-}. Columns 4--6 give the uncertainties in this prediction: column 4 is the error from measurement errors in the reference transitions; column 5 gives the error due to theoretical uncertainty in the point-nucleus finite-size effects \cite{hh}; while column 6 is the error due to neglect of higher orders than $m^3 R^2 (Z\alpha)^6$ or $m^4 R^3 (Z\alpha)^5$. Uncertainty in values for $\alpha$ and Ry give errors significantly smaller than those listed.}
\label{ahprecise}
\end{table*}

\section{\label{sec:PPEFT}PPEFT of Spin-Half Nuclei}

The classical and quantum dynamics of first-quantized spinning relativistic particles is discussed in many sources \cite{gitman1990, casalbuoni1975, barducci1976, berezin1977, brink, vecchia}. In this framework nuclear position is described by two quantum variables: a world-line position $y^\mu(s)$ (whose dynamics incorporate nuclear recoil effects) and a Grassmann (classically anti-commuting) four-vector $\xi^\mu(s)$ which incorporates spin. Grassmann variables arise because they furnish finite-dimensional representations of rotations in the quantum Hilbert space. 

With these variables the nuclear action is given by $S_{\rm pt} + S_{\rm str}$. The first term describes the interactions of an electrically charged spinning point mass,
\be \label{Sp0def}
  S_{\rm pt} = - \int   \exd s \,  \left\{ \sqrt{-\dot{y}^2}\; M + i\xi^\mu \dot{\xi}_\mu -  (Ze) \dot{y}^\mu A_\mu  \right\}  \,,
\ee
where $M$ and $Ze$ denote the mass and charge while $s$ is an arbitrary world-line parameter. The electromagnetic field $A_\mu$ is here evaluated at $x^\mu = y^\mu(s)$. $S_{\rm str}$, given below, contains all effects of nuclear substructure.

Once quantized, the Grassmann variables satisfy the anticommutation relation \cite{brink}:
\begin{equation}
\label{clifford}
\left\{ \hat{\xi}^\mu, \hat{\xi}^\nu \right\} = -\frac{1}{2}\, \eta^{\mu\nu} \,,
\end{equation}
which for relativistic spin-half nuclei are represented in terms of Dirac matrices,
\begin{equation} \label{xirep}
\hat{\xi}^\mu = \frac{i}{2} \,\Gamma^\mu \,.
\end{equation}
(We reserve lower-case $\gamma^\mu$ for matrices that act on the bulk electron field $\Psi$.)

The most general lowest-dimension interactions between $y^\mu$, $\xi^\mu$ and the electronic and electromagnetic fields relevant to current atomic measurements then are
\begin{eqnarray}\label{SpForm}
&&S_{\rm str} = \int \exd s \; \left\{  i\mu_\ssN \sqrt{-\dot{y}^2} \;\xi^\mu \xi^\nu F_{\mu\nu}  +  i c_{\rm em} \dot{y}^\mu \xi^\rho \xi^\sigma \partial_\mu F_{\rho\sigma} \right. \nn \\
&& \qquad -\ol{\Psi} \left[ \sqrt{-\dot{y}^2} \left( c_s  + i c_\ssF \xi^\mu \xi^\nu \gamma_{\mu\nu} + i c_1 \xi_5 \gamma_5 \right)   \right. \notag \\
&& \qquad \qquad  \Big. \Big. + i \dot{y}^\mu \left( c_v \gamma_\mu + c_2 \xi_5 \gamma_5 \gamma_\mu \right) \Big] \Psi  \Big\} 
\end{eqnarray}
where $\xi_5$ is shorthand for $\epsilon_{\mu \nu \rho \sigma} \xi^\mu \xi^\nu \xi^\rho \xi^\sigma$ while $\mu_\ssN$, $c_{\rm em}$, $c_s$, $c_v$, $c_\ssF$, $c_1$, $c_2$ and so on are effective couplings, with $\mu_\ssN$ having dimension (length) while the rest have dimension (length)${}^2$.

Terms in this action involving $\dot\bfy(s)$ describe nuclear-recoil effects, and are suppressed in electronic energies by inverse powers of the large nuclear mass $M$. Although some of these are required for current experimental accuracy, we temporarily drop them for the remainder of this section because they play no role in our counting of nuclear structure parameters. This allows us to specialize to the nuclear rest-frame ($\bfy = 0$), and to simplify nuclear Dirac matrices by projecting out the antiparticle sector and keeping only those terms unsuppressed by nuclear velocity:
\be \label{Gammavis0}
   \Gamma^0 \to -i\, \mathds{1}_{2\times2} \,, \quad   
   \Gamma_5 \Gamma_k \to -i \tau_k \,, \quad  
   \Gamma^{ij} \to \frac{1}{2} \epsilon^{ijk} \tau_k \,,
\ee
where $\Gamma^{\mu\nu} := -\frac{i}{4} \left[ \Gamma^\mu, \Gamma^\nu \right]$, and $\tau_k$ are $2\times 2$ Pauli matrices acting in nuclear spin-space (with the same matrices acting in electron-spin space being denoted $\sigma_k$). 


With these choices the field equations for the electron and electromagnetic fields become  
\begin{eqnarray} \label{EMptEqs}
\nabla \cdot \bfE   &=& -ie \ol{\Psi} \gamma^0 \Psi + Ze\,  \delta^3(\bfx)\,, \notag \\
(- \partial_t \bfE + \nabla \times \bfB)^i   &=& -ie \ol{\Psi} \gamma^i \Psi + \mu_\ssN \, \epsilon^{ilk} I_k \partial_l  \delta^3(\bfx)\,,
\end{eqnarray}
where $\bfI := \frac12 \, \boldsymbol{\tau}$ denotes the nuclear spin vector, and
\bea \label{Psieq}
0 &=& \left[ \gamma^0 \left( \partial_0 + i e A_0 \right) + \gamma^i \left( \partial_i + ie A_i \right) + m \right] \Psi \\
&&\quad + \delta^3(\bfx) \left[ c_s - i c_v \gamma^0 + \frac{c_\ssF}{2} \epsilon^{ijk} I_k \gamma_{ij}   \right] \Psi \,.\nn
\eea

Following standard practice, we work perturbatively in quantum-field interactions like $eA_\mu \ol{\Psi} \gamma^\mu \Psi$, whose contributions are computed using the Feynman graphs for QED, working within the interaction picture (and in Coulomb gauge). In this framework interactions between these quantum fields and the nucleus are encoded by boundary conditions obtained by integrating the above field equations over a small Gaussian sphere, $\cS_\epsilon$, of radius $\epsilon$, since these impose boundary conditions on the bulk-field mode functions at $r = \epsilon$. 

The boundary conditions found this way for the electromagnetic field are standard ones, such as the Gaussian expression for the electrostatic potential
\be \label{A0BCeq}
   4\pi \left( r^2\partial_r A_0 \right)_{r = \epsilon} =  Ze \,,
\ee
and a slightly more complicated integral (that projects onto the first spherical harmonic) for the vector potential  \cite{EFTBook, jackson, griffithsem}. In the interaction picture the solutions to the Maxwell equations subject to these boundary conditions are 
\be \label{NucEMFs}
   A^0 = A^0_{\rm nuc} = \frac{Ze}{4\pi r} , \hspace{18pt} \bfA  =  \frac{\boldsymbol{\mu} \times \bfr}{4\pi r^3} + \bfA_{\rm rad}\,,
\ee
which reveals $\boldsymbol{\mu} :=  \mu_\ssN \bfI$ to be the nuclear magnetic moment. Here $\bfA_{\rm rad}(\bfr,t)$ denotes the operator-valued radiation component of the interaction-picture electromagnetic field (whose boundary conditions are the standard, nucleus-independent ones).

The bulk Dirac equation is similarly solved in the interaction picture, with the Coulomb potential included in the unperturbed system, leading to the standard Dirac-Coulomb spinors \cite{ppeft3, ll}:
\begin{equation} \label{OmegaDef1}
\psi = \left( \begin{array}{c} \Omega_{jlj_z\varpi}(\theta,\phi) \; \mff_{nj}(r) \\ i\Omega_{jl'j_z\varpi}(\theta,\phi) \; \mfg_{nj}(r) \end{array} \right),
\end{equation}
where $\Omega_{jlj_z\varpi}$ are standard spinor spherical harmonics, whose quantum numbers $(j, l, j_z)$ are built from the mode's angular-momentum and parity quantum numbers, $(j,j_z,\varpi = \pm)$. 

The radial mode functions $\mff_{nj}(r)$ and $\mfg_{nj}(r)$ found by solving the radial part of the Dirac equation  -- see \cite{ppeftA,ppeft3}  --  with the Coulomb potential are given by  
\bea
\label{dc}
\mff_{nj}(r)  &=& \sqrt{m+\omega} \; e^{-{\rho}/{2}}  \left\{ \mathscr{C}  \rho^{\zeta-1}\left[ \mathcal{M}_1 - \left( \frac{a}{c} \right) \mathcal{M}_2 \right] \right.\nn\\
&& \quad \left. + \mathscr{D}  \rho^{-\zeta-1} \left[ \mathcal{M}_3 -  \left( \frac{a'}{c} \right) \mathcal{M}_4 \right] \right\},  
\eea
and
\bea
\label{dcg}
\mfg_{nj}(r)   &=&     - \sqrt{m-\omega} \; e^{-{\rho}/{2}} \left\{\msC \rho^{\zeta-1}  \left[ \mathcal{M}_1 + \left( \frac{a}{c} \right) \mathcal{M}_2 \right] \right. \nn\\
&& \quad \left. + \msD \rho^{-\zeta-1}  \left[ \mathcal{M}_3 +  \left( \frac{a'}{c} \right) \mathcal{M}_4 \right] \right\}, 
\eea
where $\msC$ and $\msD$ are integration constants, $\rho := 2\kappa r$ with $\kappa := \sqrt{m^2 - \omega^2}$ for mode energy $\omega$, and the functions $\mathcal{M}_\mathfrak{i}$ are given in terms of confluent hypergeometric functions
\begin{eqnarray}\label{hypergeos}
&&\!\!\!\! \mathcal{M}_1 :=   {}_1\mathcal{F}_1\left(a, b; \rho \right)\,, \; \mathcal{M}_2 :=   {}_1\mathcal{F}_1 \left( a+1, b;\rho \right)\,,\nn\\ \;\;
&&\!\!\!\! \mathcal{M}_3 :=  {}_1\mathcal{F}_1\left( a', b'; \rho\right)\,, \;  \mathcal{M}_4 :=  {}_1\mathcal{F}_1\left( a'+1, b'; \rho \right) .
\end{eqnarray}
The parameters appearing in these expressions are
\begin{eqnarray}
\label{hyperparam}
a &:=& \zeta-\frac{Z\alpha \omega}{\kappa}, \hspace{8pt} a' :=  -\left( \zeta + \frac{Z\alpha\omega}{\kappa} \right), \hspace{8pt} b := 1+ 2\zeta, \hspace{8pt}  \notag \\
b' &:=& 1-2\zeta, \hspace{8pt} c := \mfK  - \frac{Z\alpha m}{\kappa}, \hspace{6pt} \zeta := \sqrt{\mfK ^2 - (Z\alpha)^2} \,,
\end{eqnarray}
where $\mfK  := -\varpi \left( j + \frac{1}{2} \right)$.  For these solutions, only $\mathcal{M}_1$ and $\mathcal{M}_2$ are bounded at the origin, and so the radial functions are bounded there only when $\msD = 0$ (and it is this choice that leads to the standard Dirac-Coulomb energy levels for a point-like nucleus). 

The ratio $\msD/\msC$ is determined by the boundary condition for the Dirac field dictated \cite{ppeft1,ppeft2,ppeft3,ppeftA} by the $\delta$-function terms in \pref{Psieq}, which for $j = \frac12$ states imply \cite{PPEFT-Hyd}
\bea
\label{bc++}
\hat{c}_s - \hat{c}_v + X_{\ssF} \,\hat{c}_\ssF &=& \frac{ \mfg_{n \frac{1}{2} +}(\epsilon)}{ \mff_{n \frac{1}{2} +}(\epsilon)}, \\
\hat{c}_s + \hat{c}_v + X_{\ssF} \, \hat{c}_\ssF &=& \frac{\mff_{n \frac{1}{2} -}(\epsilon)}{\mfg_{n \frac{1}{2} -} \,.(\epsilon)}. \label{bc--}
\eea
Here $\hat{c}_i = {c_i}/({4\pi \epsilon^2})$ and $X_\ssF := \frac12 \left[F(F+1) - \frac32\right]$, where $F = 0,1$ is the quantum number for total atomic angular momentum, including both electron and nucleus. {\it It is only through these boundary conditions that the news about nuclear substructure gets out to the electron}, and does so through the nonzero value of $\msD/\msC$ this implies. To the accuracy needed below only the ratio $\msD/\msC$ for $j = \frac12$ modes turns out to contribute, 
\be \label{DC01}
  (\msD/\msC) = (\msD/\msC)^{(0)} + X_\ssF (\msD/\msC)^{(1)} \,,
\ee
obtained by solving \pref{bc++} and \pref{bc--} to this order. It is ultimately because all nuclear-substructure effects enter into electronic energies only through $(\msD/\msC)^{(0)}$ and $(\msD/\msC)^{(1)}$ that they can be captured entirely using two parameters.

\section{\label{sec:Comps}Predictions for Precision H Transitions}

\begin{table*}
\centering
\begin{tabular}{|c|c|c|}
\hline
Transition & $\left(2S_{1/2}^{\ssF=1} - 2S_{1/2}^{\ssF=0}\right)$ & $\left(2S_{1/2}^{\ssF=1} - 1S_{1/2}^{\ssF=1}\right)$ \\
\hline
Experimental value & $177 \, 556.834 \,3$ (kHz) & $2\, 466 \, 061 \, 102 \, 474.806$ (kHz) \\
\hline
Experimental error & $0.0067$ (kHz) & $0.010$ (kHz) \\
\hline
\hline
Pt. nucl. theory & $177 \, 564.05$ (kHz) & $2 \, 466 \, 061 \, 103 \, 430.12$ (kHz)\\
\hline
Pt. nucl. error & $0.57$ (kHz) & $0.57$ (kHz) \\
\hline
\hline
Inferred param. & $(m \epsilon_{\ssF})^2$ & $(m \epsilon_{\star})^2$ \\
\hline
Fitted value & $3.71\times 10^{-8} $  & $2.1020 \times10^{-11}$ \\
\hline
Prop. exp. error & $ 0.0035 \times 10^{-8}$  & $0.000034 \times 10^{-11}$ \\
\hline
Prop. theory error & $0.29 \times 10^{-8}$  & $0.0025 \times 10^{-11}$ \\
\hline 
\end{tabular}
\caption{The experimental values and errors (rows 2 and 3) and the point-nucleus predictions and errors (rows 4 and 5) for the reference transitions in atomic Hydrogen used for fixing the values of the two nuclear parameters listed in row 6. The last 3 rows give the values inferred for these parameters (row 7) and the errors they inherit due to the experimental uncertainty (row 8) and the precision of the point-nucleus calculation (row 9).}
\label{sat}
\end{table*}

Predicting electronic energy shifts from these formulae is straightforward (though tedious). Working to an accuracy up to and including $\delta \varepsilon \sim m^3 R^2 (Z\alpha)^6$ and $\delta \varepsilon \sim m^4 R^3 (Z\alpha)^5$ suffices to capture effects down to an accuracy of a few kHz in Hydrogen, and to this accuracy the result is $\varepsilon_{n\ssF j \varpi}  = \varepsilon_{n\ssF j \varpi}^{\rm pt} + \varepsilon_{n\ssF j \varpi}^{\NS}$ where $\varepsilon_{n\ssF j \varpi}^{\rm pt}$ is the result one would have obtained using a point-like spinning nucleus, including QED, nuclear magnetic field, and recoil corrections. Predicting $\varepsilon_{n\ssF j \varpi}^{\NS}$ is our main interest because it carries all of the contributions that explicitly involve nuclear substructure. 

\subsection{Theoretical expressions}

At the sub-kHz accuracy required here the nuclear finite-size contributions come from several sources, all with their roots in the boundary conditions \pref{bc++} and \pref{bc--} which distort the electronic mode functions by requiring a nonzero value for $\msD/\msC$. One finds
\be \label{energylevelanswer4x}
    \varepsilon_{n\ssF j \varpi}^{\NS}  =  \varepsilon_{n j \varpi}^{(0)}  + \varepsilon_{n\ssF j \varpi}^{(1)}  \,.
\ee
where $\varepsilon_{n j \varpi}^{(0)}$ contains nuclear-spin independent contributions while $\varepsilon_{n\ssF j \varpi}^{(1)}$ depends on nuclear spin proportional to the variable $X_\ssF$. In detail $\varepsilon_{n j \varpi}^{(0)} = \delta \omega_{n j \varpi}^{(0)} + \delta\varepsilon^{\QED\,(0)}_{n\ssF j\varpi}$ receives contributions both from the spin-independent part of the shift in Dirac-Coulomb mode energies due to $(\msD/\msC)^{(0)}$ of eq.~\pref{DC01} being nonzero, and from the dependence on $(\msD/\msC)^{(0)}$ appearing in QED radiative corrections, $\delta \varepsilon^{\QED\,(0)}_{n\ssF j\varpi}$. The contribution $\varepsilon_{n\ssF j \varpi}^{(1)} = \delta \omega_{n j \varpi}^{(1)} + \delta \varepsilon_{n j \varpi}^{(1)} + \delta \varepsilon^{\QED\, (1)}_{n\ssF j\varpi}$ similarly contains the spin-dependent effect of having nonzero $(\msD/\msC)^{(1)}$ in both Dirac mode energy and QED radiative corrections, plus the effects of nonzero $(\msD/\msC)^{(0)}$ in the spin-dependent energy shift due to the nuclear magnetic field ({\it i.e.}~nuclear-size corrections to hyperfine structure). These last contributions are computed perturbatively in the small parameter
\be \label{exxdef}
  \exx := \frac{m e \mu_\ssN}{4\pi} = \frac{g_p  Z \alpha\, m}{2M} \sim mRZ\alpha  \ll 1 \,,
\ee
where $g_p$ is the proton's Land\'e $g$-factor, and linear order suffices to capture nuclear-substructure effects to the order of interest. Although point-nucleus recoil corrections can be observably large, their dependence on nuclear substructure is small enough to be ignored here.

The energy shifts $\delta \omega_{n j \varpi}^{(0)}$ and $\delta \omega_{n\ssF j \varpi}^{(1)}$ are large enough to matter only for $nS_{1/2}$ and $nP_{1/2}$ (parity-even and parity-odd $j=\frac12$) states. For $nS_{1/2}$ states these work out to be \cite{PPEFT-Hyd, ppeftA}
\bea
\label{ppeftAenergy+}
&&\delta \omega_{n \frac12 +}^{(0)} = \frac{8(Z\alpha)^2m_r^3\, \epsilon_{\star}^2}{n^3}\left( \frac{m_r}{m} \right)^2  \left\{ 1 + (Z\alpha)^2 \left[ \frac{12n^2-n-9}{4n^2(n+1)} \right. \right.  \nn\\
&&\qquad \quad \left. \left. - \ln\left( \frac{2Z\alpha m_r \epsilon_{\star}}{n} \right) + 2  - \gamma - H_{n+1} \right] \right\} \,,
\eea
and
\be 
\label{spinenergy}
\delta \omega_{n \ssF \frac12 +}^{(1)} =  - \exx X_\ssF \frac{8 (Z\alpha)^2}{n^3} \left( \frac{m_r}{m} \right)^2 m_r^3 \, \epsilon_\ssF^2  \,,
\ee
while for $nP_{1/2}$ states one instead finds \cite{PPEFT-Hyd, ppeftA}
\bea
\label{ppeftAenergy-}
\delta \omega_{n \frac12 -}^{(0)} &=& 2\left( \frac{n^2-1}{n^5} \right) (Z\alpha)^4 \left( \frac{m_r}{m} \right)^2 m_r^3  \epsilon_{\star}^2, 
\eea
where $m_r = mM/(m+M)$ is the system's reduced mass and terms higher order in $Z\alpha$ are dropped. 

In these expressions $\epsilon_\star$ and $\epsilon_\ssF$ are two convenient proxies that express the nuclear-substructure dependence contained in $(\msD/\msC)^{(0)}$ and $(\msD/\msC)^{(1)}$ in a more physical way \cite{PPEFT-Hyd}. Intuition for their physical interpretation can be found by comparing eqs.~\pref{ppeftAenergy+} through \pref{ppeftAenergy-} with explicit nuclear models, which give \cite{zemach, friar, eides, pachucki2018, kalinowski2018}
\bea \label{eepsstarQED}
&&\epsilon_\star^2 = \frac{(Z\alpha)^2}{12} \left( \frac{m}{m_r}\right)^2 \\
&&\quad \times \Bigg\{ \langle r^2 \rangle_c\left[ 1 + (Z\alpha)^2 \left( 1 + \frac{1}{2}\ln \left[ \left( \frac{m}{m_r} \right)^2 \frac{(Z\alpha)^2 \langle r^2 \rangle_c}{12 \langle r_{\scriptscriptstyle C 2} \rangle^2} \right] \right)  \right. \Bigg. \notag \\
&&\qquad \Bigg. \Bigg. \qquad + \alpha(Z\alpha) \Big(4\ln2 -5 \Big) \Bigg]  -\frac{1}{2} m_r(Z\alpha) \langle r^3 \rangle_{cc}^{\rm eff}  \Bigg\},\nn
\eea
and
\be \label{eepsFQED}
\epsilon_\ssF^2 := \frac{(Z\alpha)^2 \langle r \rangle_{cm}}{4m_r}  \left\{ 1 + \frac{\alpha}{\pi} \left( \frac23 \left[ \ln \left( \frac{\Lambda^2}{m^2} \right) - \frac{317}{105} \right] - \frac54 \right)\right\} 
\ee
where the quantities $\langle r^2 \rangle_c$, $\langle r_{\scriptscriptstyle C 2} \rangle$, $\langle r^3 \rangle_{cc}^{\rm eff}$ and $\langle r \rangle_{cm}$ are various nuclear moments -- {\it e.g.}~Friar, Zemach and so on -- that arise in detailed nuclear calculations. The terms involving $\alpha$ not in the combination $Z\alpha$ express how nuclear substructure contributes to QED radiative corrections, and it is because the radiative correction involves high-energy scales that must be within a Compton wavelength of the nucleus that the effective theory allows this to be captured as if it were a nuclear moment. Besides providing an intuition for the parameters $\epsilon_\star$ and $\epsilon_\ssF$, expressions \pref{eepsstarQED} and \pref{eepsFQED} confirm our basic assertion: all of the variety of nuclear moments only appear in atomic energy shifts (at this order) through two independent parameters ($\epsilon_\star$ and $\epsilon_\ssF$). 

\subsection{Numerical results}

Given the above parameter-counting it becomes possible to make predictions for the contributions of nuclear structure to a large number of Hydrogen transitions without relying on detailed nuclear calculations (and the large nuclear uncertainties these inevitably bring). One simply uses two particularly well-measured atomic transitions to determine the two nuclear parameters from observations, and then uses these to predict the nuclear influence on all other levels. This procedure has the advantage that its errors improve with more accurate observations and with higher-precision point-nucleus calculations.

We take our reference transitions to be 
\begin{eqnarray}
\label{ah}
\nu \left( 2S^{F=1}_{\frac{1}{2}} - 2S^{F=0}_{\frac{1}{2}} \right) &=:&  177 \, 556.834 \, 3 \, (67) \hspace{6pt} \mathrm{kHz}, \nn \\
\nu \left( 2S^{F=1}_{\frac{1}{2}} - 1S^{F=1}_{\frac{1}{2}} \right) &=:&  2\,466\,061\,102\,474.806 \, (10) \hspace{6pt} \mathrm{kHz},\nn
\end{eqnarray} 
where the bracketed number indicates the size of the measurement error in the last two digits. Applying the above formulae to compute the nuclear-structure part of these transition frequencies allows an experimental determination of the two parameters $\epsilon_\star$ and $\epsilon_\ssF$, with results shown in Table \ref{sat}. 

With these values of $\epsilon_\star$ and $\epsilon_\ssF$ the nuclear-structure component of any other Hydrogen transition can be computed using eqs.~\pref{ppeftAenergy+} through \pref{ppeftAenergy-}, and Table \ref{ahprecise} provides these predictions for a list of well-measured Hydrogen transitions. Not surprisingly, our central values agree with nuclear calculations \cite{hh}, though with slightly smaller errors. Because our errors come entirely from experiments or from point-nucleus calculations, they can be expected to shrink over time in a way not limited by uncertainties in nuclear models. Details of these calculations and predictions for a much longer list of transitions can be found in \cite{PPEFT-Hyd}. 

\section{\label{sec:conc}Conclusion}

In summary, we collect and summarize the results of \cite{PPEFT-Hyd}, which uses recently developed tools \cite{ppeft1, ppeft2, ppeft3, ppeftA} to compute how nuclear-substructure can modify atomic energy levels up to (and including) contributions of order $m^3 R^2 (Z\alpha)^6$ and $m^4 R^3 (Z\alpha)^5$ for atomic Hydrogen. We show that all short-distance effects (which include the Friar, Zemach and related moments, size-dependent corrections to nuclear polarizability and QED radiative processes) enter as contributions to only two independent parameters, $\epsilon_\star$ and $\epsilon_\ssF$. Of these, $\epsilon_\ssF$ captures all of the effects that depend on nuclear spin (and so is not present for spinless nuclei). The parameter $\epsilon_\star$ captures all spin-independent nuclear effects and so is the only relevant parameter for spinless nuclei, as found earlier in \cite{ppeftA}. Ref.~\cite{PPEFT-Hyd} explicitly verifies that extant nuclear calculations do indeed contribute to atomic energies through these two parameters (as they must).

This counting of parameters is extremely robust because it relies only on general principles of effective field theories (EFTs) and not at all on nuclear modelling. Indeed it applies equally well to {\it any} short-distance physics localized near the nucleus, and this is the underlying reason why the same parameters also capture how nuclear structure enters into QED radiative corrections. The main novel contribution of the EFT methods used is to solve the technical issue concerning near-nucleus divergences that generically arise in the electron's Dirac mode functions, due to the singular behaviour they develop near the origin due to the nuclear presence. This divergent behaviour is a calculational artefact because in reality nuclear structure intervenes to cut it off, but we show in \cite{PPEFT-Hyd} that they can instead be renormalized into the effective nuclear couplings within an EFT framework. 

We have used the knowledge that nuclear physics is limited to two parameters to compute nuclear effects for a variety of transitions in atomic Hydrogen. Numerically, the two independent parameters $\epsilon_\star$ and $\epsilon_\ssF$ control all nuclear contributions down to order $10^{-2}$ kHz in atomic Hydrogen. By fitting these two parameters to two particularly well-measured transitions, we predict the nuclear-size contributions to a number of well-measured Hydrogen transitions, providing nuclear predictions for these transition energies that are not limited by the uncertainties intrinsic to nuclear models. 

A similar exercise also works for muonic Hydrogen, although with different numerical parameters. Two parameters in this case suffice to capture all nuclear effects down to errors of order 0.01 meV. Because of the larger muon mass, contributions of order $m^5 R^4 (Z\alpha)^6$ are also required to match the accuracy of current observations, and these go beyond the analysis done here. These techniques generalize to other Hydrogen-like nuclei with arbitrary spin, and should be of most value for those with the largest internal gap for exciting internal nuclear degrees of freedom.

\begin{acknowledgments}
We thank Eric Hessels, Marko Horbatsch, Sasha Penin, Randolph Pohl, Ira Rothstein and Kai Zuber for helpful discussions while developing these ideas.  We thank the organizers of the workshop `Precision Measurements and Fundamental Physics: the Proton Radius and Beyond' held at the Mainz Institute for Theoretical Physics (MITP), for providing such stimulating environs where part of this work was completed.
This research was supported in part by funds from the Natural Sciences and Engineering Research Council (NSERC) of Canada. Research at the Perimeter Institute is supported in part by the Government of Canada through Industry Canada, and by the Province of Ontario through the Ministry of Research and Information (MRI). 
\end{acknowledgments}


\begin{thebibliography}{99}

\bibitem{PPEFT-Hyd}
L.~Zalavari, C.~P.~Burgess, P.~ Hayman and M.~Rummel,
``Precision Nuclear-Spin Effects in Atoms: EFT Methods for Reducing Theory Errors'',
 arXiv:2008.09718 [hep-ph].

\bibitem{ppeftA}
 C.~P.~Burgess, P.~Hayman, M.~Rummel and L.~Zalavari,
  ``Reduced theoretical error for $^4He^+$ spectroscopy,''
  Phys.\ Rev.\ A {\bf 98} (2018) no.5,  052510
  [arXiv:1708.09768 [hep-ph]].
 
 \bibitem{boundNRQED} 
  P.~Labelle,
  ``Effective field theories for QED bound states: Extending nonrelativistic QED to study retardation effects,''
  Phys.\ Rev.\ D {\bf 58}, 093013 (1998)
  [hep-ph/9608491].

\bibitem{ppeft1}
 C.~P.~Burgess, P.~Hayman, M.~Williams and L.~Zalavari,
  ``Point-Particle Effective Field Theory I: Classical Renormalization and the Inverse-Square Potential,''
  JHEP {\bf 1704} (2017) 106
  [arXiv:1612.07313 [hep-ph]].
  
\bibitem{ppeft2}
 C.~P.~Burgess, P.~Hayman, M.~Rummel, M.~Williams and L.~Zalavari,
  ``Point-Particle Effective Field Theory II: Relativistic Effects and Coulomb/Inverse-Square Competition,''
  JHEP {\bf 1707} (2017) 072
  [arXiv:1612.07334 [hep-ph]].
  
\bibitem{ppeft3}
C.~P.~Burgess, P.~Hayman, M.~Rummel and L.~Zalavari,
  ``Point-Particle Effective Field Theory III: Relativistic Fermions and the Dirac Equation,''
  JHEP {\bf 1709} (2017) 007
  [arXiv:1706.01063 [hep-ph]].

\bibitem{falltocenter}
R.~Plestid, C.~P.~Burgess and D.~H.~J.~O'Dell,
 ``Fall to the Centre in Atom Traps and Point-Particle EFT for Absorptive Systems,''
 JHEP {\bf 1808} (2018) 059
 [arXiv:1804.10324 [hep-ph]].


\bibitem{friar}
J.~L.~Friar,
``Nuclear finite-size effects in light muonic atoms'',
Ann. Phys. {\bf 122} (1978) 152-196
  
\bibitem{zemach}
A.~C.~Zemach,
  ``Proton Structure and the Hyperfine Shift in Hydrogen,''
  Phys.\ Rev.\  {\bf 104} (1956) 1771.
  
\bibitem{eides}  
 M.~I.~Eides, H.~Grotch and V.~A.~Shelyuto,
  ``Theory of Light Hydrogenic Bound States,''
  Springer Tracts Mod.\ Phys.\  {\bf 222} (2007) pp. 1.
    
\bibitem{pachucki2018}
  K.~Pachucki, V.~Patk\'o\v  s and V.~A.~Yerokhin,
  ``Three-photon exchange nuclear structure correction in hydrogenic systems,''
  Phys.\ Rev.\ A {\bf 97} (2018) no.6,  062511
  [arXiv:1803.10313 [physics.atom-ph]].
  
\bibitem{kalinowski2018}
  M.~Kalinowski, K.~Pachucki and V.~A.~Yerokhin,
  ``Nuclear-structure corrections to the hyperfine splitting in muonic deuterium,''
  Phys.\ Rev.\ A {\bf 98} (2018) no.6,  062513
  [arXiv:1810.06601 [physics.atom-ph]].
  
\bibitem{hessels2019}
  N.~Bezginov, T.~Valdez, M.~Horbatsch, A.~Marsman, A.~C.~Vutha and E.~A.~Hessels,
  ``A measurement of the atomic Hydrogen Lamb shift and the proton charge radius,''
  Science {\bf 365} (2019) no.6457,  1007.

\bibitem{kramida}
A.~E.~Kramida,
``A Critical Compilation of Experimental Data on Spectral Lines and Energy Levels of Hydrogen, Deuterium and Tritium'',
At.~Data Nucl.~Data Tables, {\bf 96}, (2010) 586-644,

\bibitem{fleurbaey}
H.~Fleurbaey, S.~Galtier, S.~Thomas, M.~Bonnaud, L.~Julien, F.~Biraben, F.~Nez, M.~Abgrall and J.~Guéna,
``New Measurement of the $1S-3S$ Transition Frequency of Hydrogen: Contribution to the Proton Charge Radius Puzzle,''
Phys. Rev. Lett. \textbf{120} (2018) no.18, 183001
[arXiv:1801.08816 [physics.atom-ph]].

\bibitem{hh} %
M.~Horbatsch, E.~A.~Hessels,
``Tabulation of the bound-state energies of atomic Hydrogen'',
Phys. Rev. A {\bf 93} (2016) 022513,

\bibitem{gitman1990}
  D.~M.~Gitman and I.~V.~Tyutin,
  ``Quantization of Fields with Constraints,''
  Springer (1990), 291 p.

\bibitem{casalbuoni1975}
  R.~Casalbuoni,
  ``Relativity and Supersymmetries,''
  Phys.\ Lett.\  {\bf 62B} (1976) 49.
  
\bibitem{barducci1976}
  A.~Barducci, R.~Casalbuoni and L.~Lusanna,
  ``Supersymmetries and the Pseudoclassical Relativistic electron,''
  Nuovo Cim.\ A {\bf 35} (1976) 377.
  
\bibitem{berezin1977}
  F.~A.~Berezin and M.~S.~Marinov,
  ``Particle Spin Dynamics as the Grassmann Variant of Classical Mechanics,''
  Annals Phys.\  {\bf 104} (1977) 336.

\bibitem{brink}
L.~Brink, P.~Di Vecchia, P.~Howe,
``A Lagrangian formulation of the classical and quantum dynamics of spinning particles'',
Nucl. Phys. B {\bf 118} (1977) 76-94

\bibitem{vecchia}
P.~Di Vecchia, F.~Ravndal,
``Supersymmetric Dirac Particles'',
Phys.\ Lett.\ A {\bf 73} (1979) 371.

\bibitem{EFTBook}
  C.P.~Burgess,
  {\it Introduction to Effective Field Theory: Thinking effectively about hierarchies of scale}, Cambridge University Press 2020 (in press).

\bibitem{jackson}
J.~D.~Jackson,
``Classical Electrodynamics'',
New York: Wiley, 2nd ed. (1975) 848 p.

\bibitem{griffithsem}
D.~J.~Griffiths,
``Introduction to electrodynamics'',
Prentice Hall, 3rd ed. (1999) 576 p.

\bibitem{ll}
V.~B.~Berestetskii, E.~M.~Lifshitz and L.~P.~Pitaevskii,
``Relativistic quantum theory'',
Pergamon Press (1971-74) vol. I.

\end{thebibliography}
\end{document}